\documentclass[a4paper,fleqn,usenatbib]{mnras}

\usepackage{newtxtext,newtxmath}

\usepackage[T1]{fontenc}
\usepackage{ae,aecompl}

\usepackage{graphicx}	
\usepackage{amsmath}	
\usepackage{amssymb}	
\usepackage{bm}

 \usepackage{subfigmat}	      	

\usepackage{color}

\newcommand{\Frac}[2]{\mbox{$\displaystyle\frac{#1}{#2}$}}
\newcommand{\Der}[2]{\Frac{d#1}{d#2}}
\newcommand{\Dernth}[3]{\Frac{d^{#3}#1}{d#2^{#3}}}

\newcommand{\Sum}{\displaystyle\sum}

\title[The solution of the Generalized Kepler's equation]{The solution of the Generalized Kepler's equation}

\author[R. L\'opez et al.]{
Rosario L\'opez,$^{1,2}$
Denis Hautesserres$^{3}$
and Juan F\'elix San-Juan$^{1}$\thanks{E-mail: juanfelix.sanjuan@unirioja.es}
\\
$^{1}$Scientific Computing Group (GRUCACI), University of La Rioja, \\26006 Logro\~no, Spain\\
$^{2}$Center for Biomedical Research of La Rioja,  26006 Logro\~no, Spain\\
$^{3}$Centre National d'\'Etudes Spatiales, 31401 Toulouse Cedex 9, France
}

\date{Accepted XXX. Received YYY; in original form ZZZ}

\pubyear{2017}

\begin{document}
\label{firstpage}
\pagerange{\pageref{firstpage}--\pageref{lastpage}}
\maketitle

\begin{abstract}
In the context of general perturbation theories,  the \textit{main problem} of the artificial satellite  analyses the motion of an orbiter around  an Earth-like planet, only perturbed by its equatorial bulge or $J_2$ effect.  By means of  a Lie transform and   the Krylov-Bogoliubov-Mitropolsky method, a  first-order theory in  closed form of the eccentricity is produced. During the evaluation of the theory it is necessary to solve a generalization of the classical Kepler's equation. In this work,  the application of a numerical technique and three initial guesses  to  the Generalized Kepler's equation are discussed.
\end{abstract}




\section{Introduction}

Kepler's equation has been studied for more than three centuries  due to its relevance in  the Celestial Mechanics and Astrodynamics fields \cite{col1993_kepeq3cent}. During this time, several different approaches have been proposed to solve this transcendental equation. Some of them are based on graphical  \cite{see1895kep_graph},  mechanical \cite{plu1906kep_mech}, analytical \cite{dep1979_lagrangeinv,lyn2015kep_ana} and numerical solutions
\cite{smi1979_startvalkepeq,ng1979_algkepeq,dan1983_kepeq1,ode1986_prockepeq,dan1987_kepeq3,taf1989_solvkepeq,nij1991_kepeqeff,fuk1996_fastkepeq,pal2002_kepeqaccnewt,rap2017kep_num}.

However, differently from the two-body problem, the  gravity field  of the planet is the main effect that disturbs the trajectory of an artificial satellite  or space debris object, so that, in general,  the two-body dynamics does not constitute  a good approximation to the true dynamics of the orbiter. In the context of the artificial satellite problem, general perturbation theories are used to provide a fast approach to the calculation  of the position and velocity of the satellite. 
 
This paper deals with a transcendental equation which generalizes  Kepler's equation. This equation appears when the Krylov-Bogoliubov-Mitropolsky method \cite{kry1943_nonlinmech} is used  so as to obtain a closed-form approximate analytical solution to the zonal satellite problem \cite{cab1975_ord2hill_phd,cal1971_promed_phd,san1994gru_atesat_cnes,san2000gru_mars_cnes,aba2001gru_shortevol,san2011gru_ppkbz9}.  The  simplest case where  this transcendental equation appears is the \textit{main problem} of the artificial satellite. This new equation, like Kepler's equation, cannot be directly inverted in terms  of simple  functions because it is transcendental, so it is usually solved through numerical methods.  It is worth noting that inaccuracy in the solution of the Generalized  Kepler's  equation  introduces an accuracy problem in the determination of the position and velocity of the satellite.

In this paper, we apply the iterative  method  proposed by Danby and Burkardt, as well as two typical initial guesses that \cite{dan1983_kepeq1} used to solve Kepler's equation. We also propose the use of the solution  of  Kepler's equation itself as an initial guess for the iterative resolution of the Generalized  Kepler's  equation, in an effort to find a method which is both simple and efficient.

\section{Kepler's Equation}\label{sec_keplereq}

Kepler's equation (KE)  relates the position of a satellite in its orbit to the time. This relationship can be expressed by the transcendental equation in $E$:

\[M = E - e \sin E,\]

\noindent
where $M$ and  $E$ are  the  mean and eccentric anomalies, respectively,  and $e$ is the  eccentricity of the orbit. The mean anomaly  is  related to the time according to:

\[M = n (t-T),  \]

\noindent
where $n$ is the mean motion, which represents the average angular velocity, that is, $2\pi$ divided by the keplerian period, and $T$ is the time of the perigee passage.

The solution to  KE  in the elliptic case, $0\leq e < 1$, consists in finding the root of the function

\begin{eqnarray}\label{equ0}
F(E) & = & E - e \sin E  - M = 0,
\end{eqnarray}

\noindent
by giving a pair of values to $e$ and $M$, where $E, M \in [0, \pi]$.  It is well known that, for other ranges of $E$ and $M$, the solutions can be obtained by simply replacing $E$, $M$ for either $-E$, $-M$ or  $E\pm 2k\pi$, $M\pm 2k\pi$, with $k$ being an integer.

Unfortunately,  inverting  KE, that is, finding the eccentric anomaly as a function of the mean anomaly and the eccentricity,  is not an easy task. In practice, iterative methods \cite{tra1982_itermeth} provide  approximate solutions to  this problem. Some of the most popular  iterative methods used to solve Eq. \eqref{equ0} are Newton-Raphson, Halley, and the one devised by Danby and Burkardt \cite{dan1983_kepeq1},  which  is known   as the Danby method in scientific literature. The iterations corresponding to these methods can be  defined as
 
\[x_{n+1}= x_n -\frac{F( x_n)}{F'( x_n)},\]

\noindent
for the Newton-Raphson method,

\[x_{n+1}= x_n -\frac{2F( x_n)F'( x_n)}{2[F'( x_n)]^2-F( x_n)F'( x_n)},\]

\noindent
for the Halley method and, finally,

\[x_{n+1}= x_n -\delta_{n3},\]

\noindent
where 

\begin{eqnarray*}
\delta_{n1}  & = & \frac{F}{F'},\\
\delta_{n2}  & = & -\frac{F}{F'+\frac{1}{2} \delta_{n1}F''},\\
\delta_{n3}  & = & -\frac{F}{F'+\frac{1}{2} \delta_{n2}F''+\frac{1}{6} \delta_{n2}^2F'''},
\end{eqnarray*}

\noindent
for the Danby method (DM), which have  quadratic, cubic and quartic convergence, respectively.

\section{First-order analytical theory}
\begin{figure}
\includegraphics[width=\columnwidth]{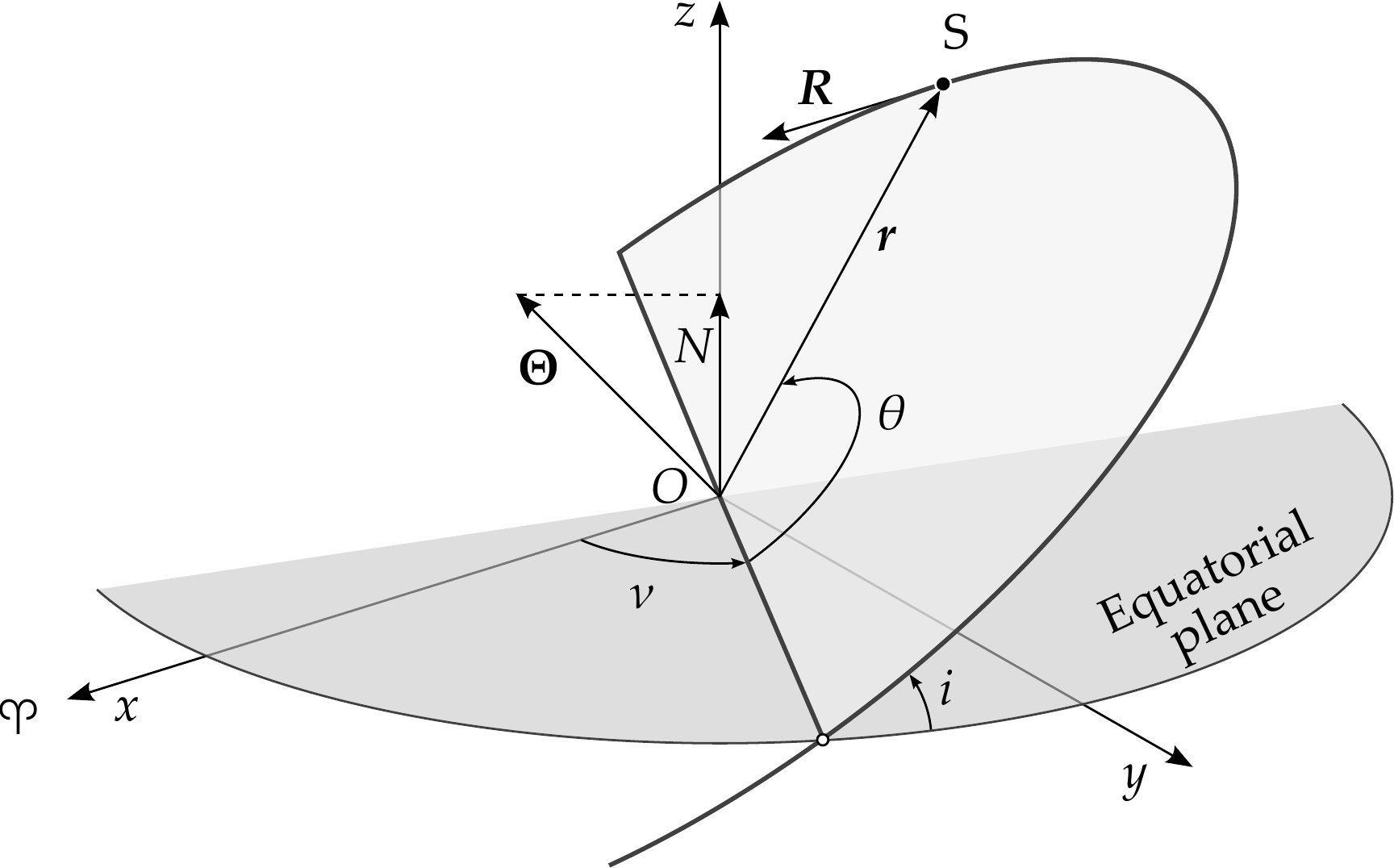}
\caption{Polar-nodal variables $(r,\theta,\nu,R,\Theta,N)$. $r$ is  the radial distance from the centre of mass of the planet to the satellite, $\theta$ is the argument of the latitude, $\nu$ represents the argument of the node, $R$ is the magnitude of the radial velocity, $\Theta$ is the magnitude of the angular momentum vector, whereas $N=\Theta \cos i$. }
\label{fig:fig1}
\end{figure}

In this section, the polar-nodal variables $(r,\theta,\nu,R,\Theta,N)$ will be used to describe  the \textit{main problem} of the artificial satellite theory. The meaning of these variables is shown in Fig. \ref{fig:fig1}. O$xyz$ represents an inertial reference frame centred at the centre of mass of the Earth-like planet. The variable $r$ denotes the distance from the centre of mass of the Earth-like planet to the satellite, $\theta$ is the argument of the latitude of the satellite, $\nu$ represents the argument of the node, $R$ is the radial velocity, $\Theta$ designates the magnitude of the angular momentum vector $\bm{\Theta}$, whereas $N$ represents the projection of $\bm{\Theta}$ onto the $z$-axis.

The \textit{main problem} of the artificial satellite theory is given by the Hamiltonian 

\begin{equation}\label{equ1}
\mathcal{H} = \mathcal{H}_\mathcal{K} + \mathcal{H}_{J_2},
\end{equation}

\noindent
where  $\mathcal{H}_\mathcal{K}$ corresponds to the Kepler problem  and $\mathcal{H}_{J_2}$ to the influence of  $J_2$, which is a positive constant representing the shape of the Earth-like planet. These terms, expressed in the polar-nodal variables, are:

\begin{eqnarray*}
\mathcal{H}_\mathcal{K} & = & \Frac{1}{2} \left( R^2 + \Frac{ \Theta^2}{r^2}\right) - \Frac{\mu}{r},\\[.5ex]
\mathcal{H}_{J_2} & = & J_{2} \Frac{\mu}{r}   \left(\Frac{\alpha}{r} \right)^{2}  P_{2} (s  \,\sin \theta).
\end{eqnarray*}

\noindent
$P_2$ is the Legendre polynomial of degree $2$, $\mu$ is the gravitational constant of the Earth-like planet, $\alpha$ is its equatorial radius and  $s$ is the sine of the inclination $i$.

This two-degree-of-freedom problem (2-DOF) is   non-integrable  \cite{iri1993_integraj2}. However, by applying perturbation theories, approximate analytical solutions can be obtained \cite{koz1962_2ordnodrag,bro1959_astnodrag}.  Considering $J_{2}$ as a small parameter $\epsilon$, Eq. \eqref{equ1} can be rewritten in the form of a perturbed Hamiltonian

\[\mathcal{H} = \mathcal{H}_0 + \epsilon \mathcal{H}_1,\]

\noindent
where $\mathcal{H}_0 = \mathcal{H}_\mathcal{K}$ and $\mathcal{H}_1=\mathcal{H}_{J_2}/J_2$.

The  perturbation theory is based on the assumption that the difference between   
$\mathcal{H}$ and $\mathcal{H}_0$ is small. Then, using the Lie transform technique, an approximate first-order closed-form analytical  solution for the \textit{main problem}  can be developed.  The elimination of the Parallax \cite{dep1981_parallax} is a Lie transform, 
$(r,\theta,\nu,R,\Theta,N) \longrightarrow (r',\theta',\nu',R',\Theta',N')$, which  removes the long-period terms, produced by the argument of the perigee,  from the transformed Hamiltonian $\mathcal{K}$, whereas the short-period terms, caused by the mean anomaly $M$, still remain in $\mathcal{K}$ through the variables $(r,R)$. It must be noted that the  argument of the latitude is the sum of the argument of the perigee and the true anomaly $f$, which  is related to the mean anomaly $l$ through Kepler's equation. Finally, the transformed Hamiltonian and the generating function of the corresponding Lie transform can be  simultaneously obtained. The expression of $\mathcal{K}$ yields
 
\[\mathcal{K} =\mathcal{K}_0 + \epsilon  \mathcal{K}_1,\]

\noindent 
where

\begin{eqnarray*}
\mathcal{K}_0 & = & \Frac{1}{2} \left( R'^2 + \Frac{ \Theta'^2}{r'^2}\right) - \Frac{\mu}{r'},\\[.5ex]
 \mathcal{K}_1 & = &  \Frac{\alpha^2 \mu^2}{\Theta'^2} \left(\Frac{1}{2} -
\Frac{3}{4} s'^2  \right) \Frac{1}{r'^2}.
\end{eqnarray*}

\noindent 
As can be observed, the argument of the latitude $\theta'$ does not  appear in the transformed Hamiltonian, which implies that    the  number of degrees
of freedom is reduced to one and, therefore, it is trivially integrable. The direct and inverse transformations can be calculated from the generating function (see Appendix A).
 
Then, $\mathcal{K}$  is transformed into a perturbed  harmonic oscillator by replacing the variables $r'$, $dr'/dt$ with two new 
variables $u,v$, respectively, and  the time $t$ with a new independent variable $\tau$:

\begin{equation}\label{equ2}
u =\Frac{1}{r'} - \frac{1}{p'},\qquad r'^2 \Der{\tau}{t} = \Theta',\qquad v = \Der{u}{\tau},
\end{equation}

\noindent
with $p'=\Theta'^2/\mu$.  Finally,  we obtain

\[\Dernth{u}{\tau}{2} + u =  \epsilon \mathcal{P} \left(\frac{1}{p'} - u\right),\]

\noindent 
where
\begin{eqnarray*}
\mathcal{P} & = & \-\Frac{\alpha^2 }{p'^2} \left(1 - \Frac{3}{2} s'^2 \right)
\end{eqnarray*}

\noindent 
is a constant.\par

Then, the Krylov-Bogoliubov-Mitropolsky method  is applied to the integration of this harmonic oscillator.  This method assumes an asymptotic expansion of the solution in the form

\[u = \delta \cos \psi + \Sum_{i\ge1} \Frac{\epsilon^i}{i!} u_i(\delta,\psi),\]

\noindent where $u_i$ are $2 \pi$-periodic functions in $\psi$, and the 
relation of $\delta$ and $\psi$ with  the fictitious time $\tau$ is given by

\begin{eqnarray*}
\Der{\delta}{\tau}  & = & \Sum_{i\ge0} \Frac{\epsilon^i}{i!} A_n(\delta) , \\[.5ex]
\quad \Der{\psi}{\tau} & = & \Sum_{i\ge0} \Frac{\epsilon^i}{i!} B_n(\delta).
\end{eqnarray*}

The values of the first order of $u$ and $v$ are provided by the Krylov-Bogoliubov-Mitropolsky  method, together with
the variation of the amplitude $\delta$ and the perturbed true anomaly $\psi$ with respect to the fictitious time $\tau$:

\begin{eqnarray*}\label{dfi}
\Der{\delta}{\tau} &=& 0,\\[.5ex]
\Der{\psi}{\tau} &=&  \nonumber
1 - \frac{ \epsilon}{2 } \mathcal{P}. 
\nonumber
\end{eqnarray*}

Finally, the expressions of the polar nodal variables are:

\begin{eqnarray*}
\Frac{p'}{r'}  &=& 1+\delta\,p' \cos \psi + \epsilon \mathcal{P},\\[.5ex]
(\theta'-\theta'_0) \Der{\psi}{\tau}  & = & \psi  + \frac{\epsilon}{2}\left(\frac{ \alpha }{ p'} \right)^2 
\left(5-6 s'^2\right) \psi,
 \nonumber\\[.5ex]
(\nu'-\nu'_0) \Der{\psi}{\tau}  & = &-\frac{ 3\,\epsilon}{2 } \left(\frac{ \alpha }{ p'} \right)^2
c' \psi,
  \nonumber\\[.5ex]
 \frac{R'}{ \delta \Theta'} &=&  \sin \psi 
+ \frac{ \epsilon}{2 }  \mathcal{P} \sin \psi, \nonumber\\[.5ex]
\Theta' & = & \Theta' _0,\nonumber\\[.0ex] 
N'& = &  N'_0 ,\nonumber
\end{eqnarray*}

\noindent
where $c'$ is the cosine of $i'$.

Combining the relations of $\psi$ with $\tau$, and $\tau$ with $t$, we obtain 
the following  relation between $\psi$ and $t$:

\begin{equation}\label{eq1}
r^2 d \psi = \Theta\Der{\psi}{\tau}  d t.
\end{equation}

In order to integrate Eq. \eqref{eq1}, an auxiliary variable $E_k$, which has a similar meaning  as the eccentric anomaly in the elliptical motion, is 
defined by the relations

\[\cos \psi = \Frac{\sqrt{1-e_k^2} \sin E_k}{1 - e_k \cos E_k},\qquad
\sin \psi= \Frac{\cos E_k - e_k}{1 - e_k \cos E_k},\]
\noindent
where $e_k=\delta p'$. After that, taking into account the relation between $u$ and $r'$ given in Eq. \eqref{equ2},  we obtain

\begin{eqnarray}\label{equ3}
M_k & = & E_k - e_k \sin E_k  
+  \epsilon  \frac{\alpha^2( 3 s^2-2)}{ 4a_k^2\left(1-e_k^2\right)^3} \left[2 \left(e_k^2+2\right) E_k 
\right. \\
&& \left. - 8 e_k \sin E_k +e_k^2 \sin 2 E_k\right],\nonumber
\end{eqnarray}

\noindent
with

\begin{equation}
M_k = \frac{\Theta}{a_k \eta_k} \Der{\psi}{\tau} (t-T), 
\end{equation}

\noindent 
where $\eta_k=\sqrt{1-e_k^2}$, $a_k=p'/\eta_k$,  and  the time $T$  corresponds to the instant when $\psi=0$  (see Reference \cite{san2009gru_ppkbz9_aas} for more details). This equation  can be considered  a perturbed case of the classical Kepler's equation, and it plays the same  role  in the  accuracy determination of the position of the satellite.

 It is worth noting that the values of $e_k$ and $a_k$ are close to the values  of eccentricity and   semi-major axis of the orbit, respectively.  That is the reason why  $e_k$ and $a_k$ will be approximated by the real values of   eccentricity and   semi-major axis. This assumption will be extended to the new anomalies  $E_k$ and $M_k$, and their behaviour compared with the eccentric and mean anomalies of the orbit, $E$ and $M$, in the   theoretical study of  Eq. \eqref{equ3} that is presented in this work. Hereinafter, with a slight  notation abuse,  we will refer to the generalized eccentric and mean anomalies with the symbols $E$ and $M$.

\section{Generalized Kepler's equation}
The first-order generalized Kepler's equation (GKE) is given by

\begin{eqnarray*}
M & = & E - e \sin E    \\
&&+ \frac{\epsilon^*}{ \left(1-e^2\right)^3} \left[2 \left(e^2+2\right) E - 8 e \sin E +e^2 \sin 2 E\right] ,\noindent
\end{eqnarray*}
   
\noindent 
where $\epsilon^*$ represents a new small dimensionless parameter which depends on the physical constants, $\epsilon=J_2$ and $\alpha$, and the generalized inclination and semi-major axis:

\begin{equation}\label{para}
 \epsilon^*= \epsilon  \left(\frac{ \alpha}{2 a }\right)^2 \left(3 s^2-2\right).
 \end{equation}

\begin{figure}
\begin{subfigmatrix}{1}
\subfigure[$\epsilon^*(a,i)$.]{\includegraphics{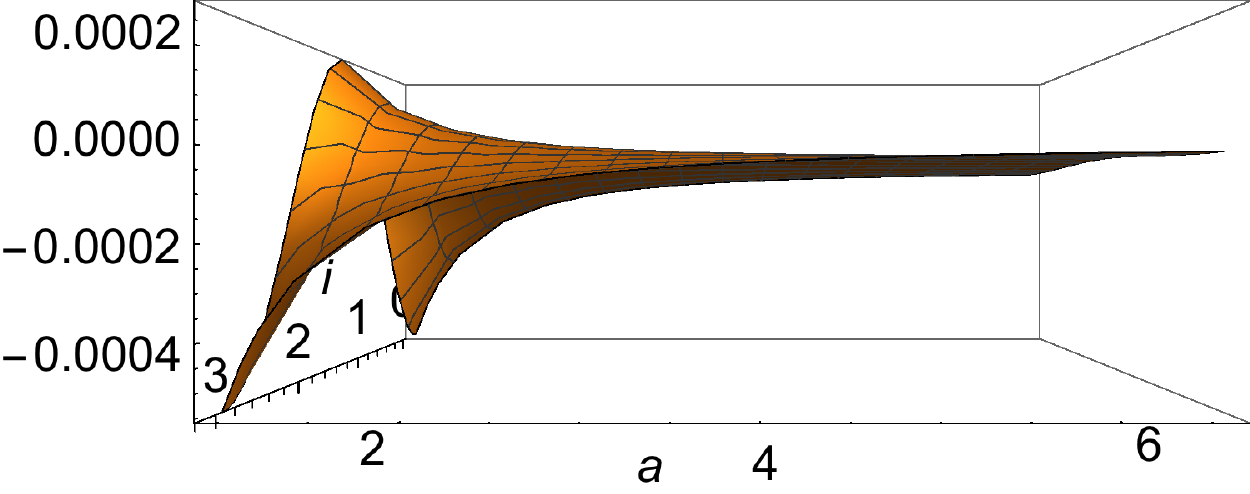}}
\subfigure[$\epsilon^*(a=\alpha,i)$. ]{\includegraphics{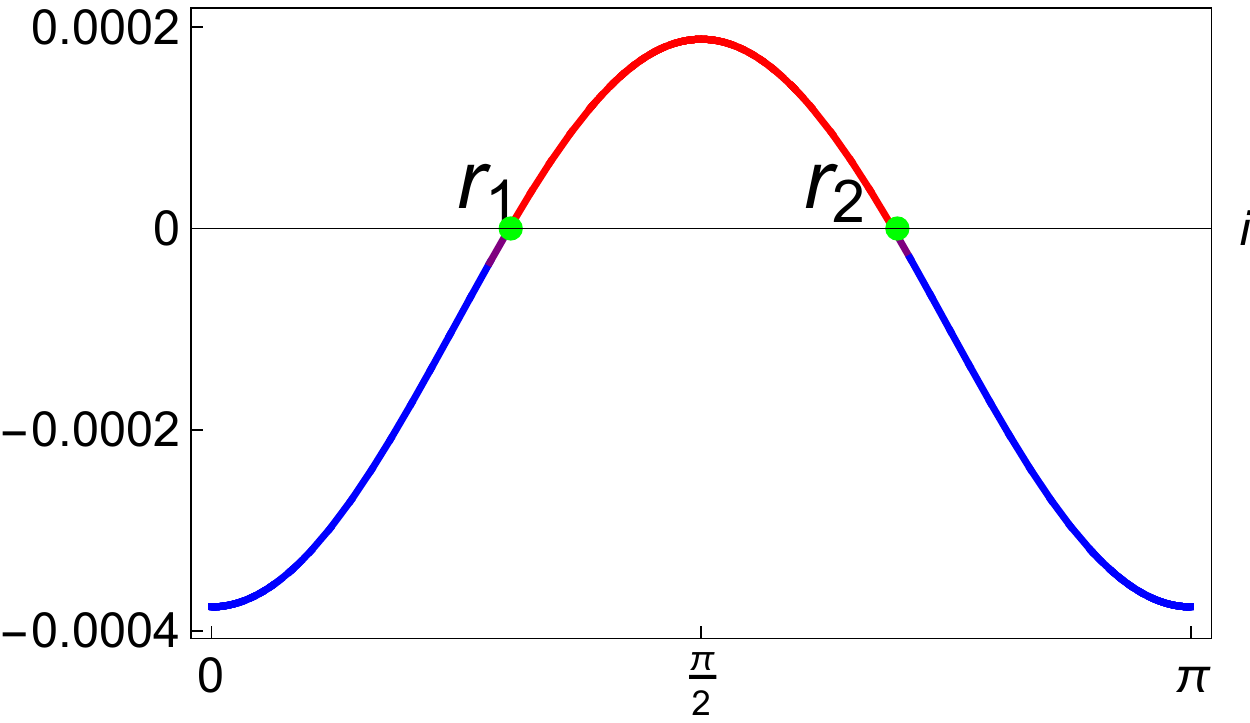}}
 \end{subfigmatrix}
\caption{Graphical representation of $\epsilon^*$. The magnitude  $|\epsilon^*|$ depends on  $a$, whereas its sign is a function of $i$.  In the case of the Earth, $J_2=0.001082626836196$, and therefore $\epsilon^* \in [-0.00054131341,0.000270656709]$.} \label{fig:fig2}
\end{figure}

\noindent
Fig. \ref{fig:fig2} (a) shows a graphical representation of $\epsilon^*(a,i)$. The units of $a$ and $i$ are   mean equatorial planet radii and radian, respectively.   The sign of $\epsilon^*$ depends on the inclination: $\epsilon^*$  takes positive values for $i\in(r_1,r_2)$,  reaching its maximum, $J_2/4$, when $i=\pi/2$ and $a=\alpha$, negative values for $i\in[0,r_1)$ and $i\in(r_2,\pi]$,  reaching its minimum, $-J_2/2$, when $i=0$  and $a=\alpha$,  and zero values for the inclinations $r_1=\arcsin(\sqrt{2/3})$ and $r_2=\pi -\arcsin(\sqrt{2/3})$, that is, the roots  of the equation $\epsilon^*=0$. For the values $r_1$ and $r_2$, the classical KE is recovered. On the other hand, the value of $|\epsilon^*|$ decreases when  $a$ increases.  Fig. \ref{fig:fig2}  (b) shows the plot of $\epsilon^*$ when the  semi-major axis takes the value of $\alpha$ $(-J_2/2\leq \epsilon^*\leq J_2/4)$.  Positive values of $\epsilon^*$ are plotted in red while negative values are in blue.

\begin{figure}
\begin{subfigmatrix}{1}
\subfigure[$e_p(a,i)$.]{\includegraphics{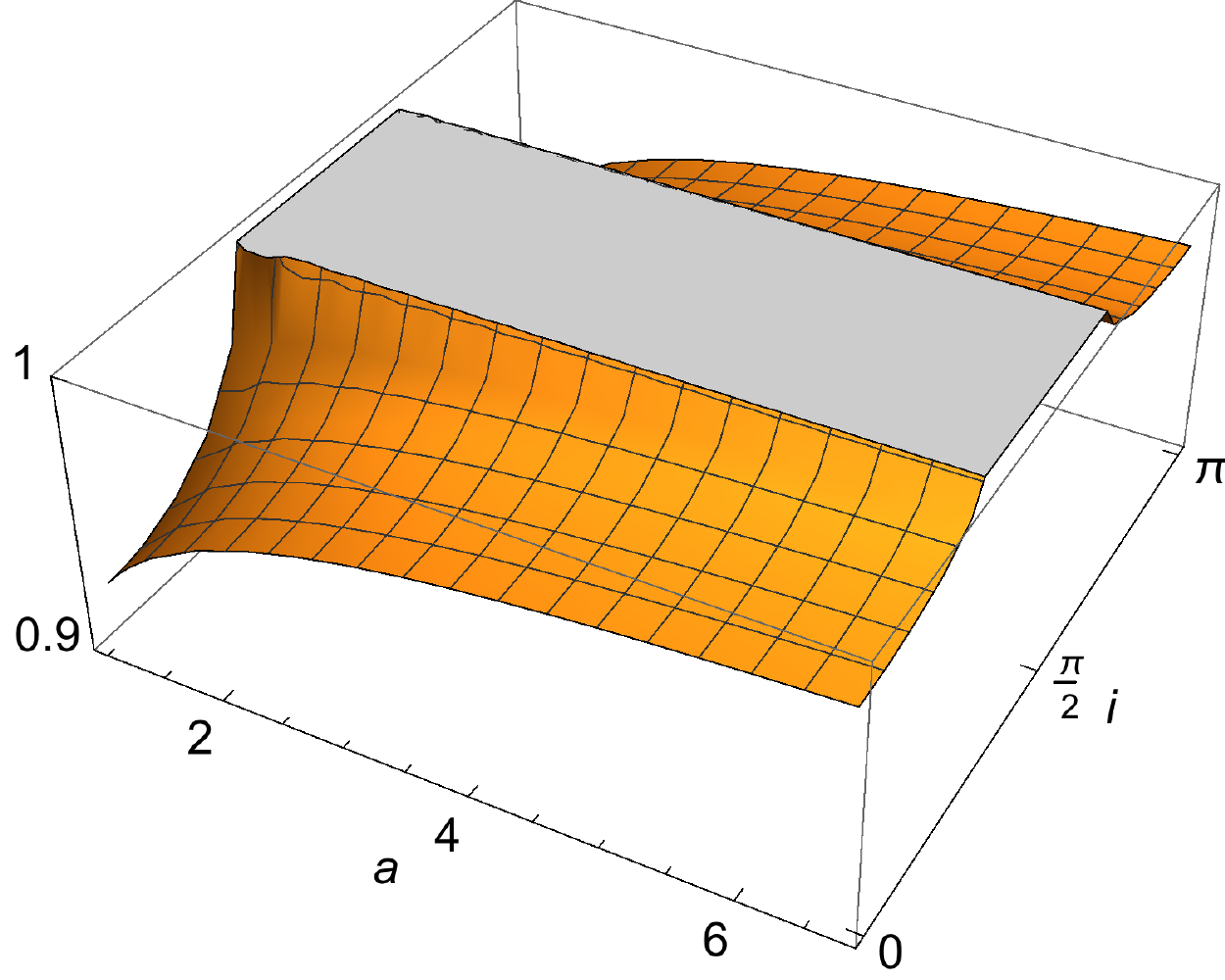}}
\subfigure[$e_p(a=\alpha,i)$. ]{\includegraphics{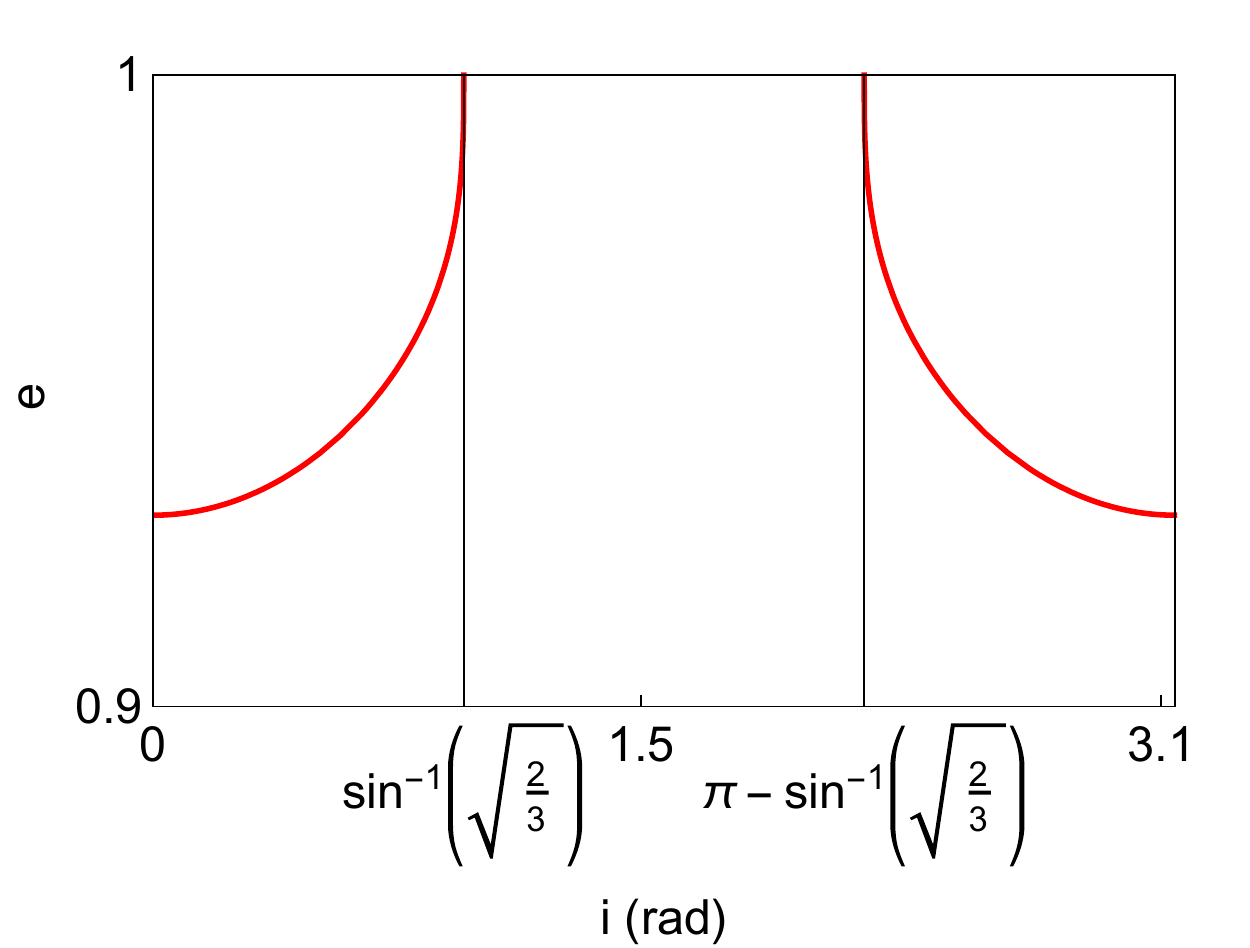}}
 \end{subfigmatrix}
\caption{Graphical representation of $e_p$. The value  $e_p$ depends on  $a$, whereas its existence is a function of $i$. }
 \label{fig:fig4}
\end{figure}

Solving the perturbed Kepler's equation in the elliptic case is equivalent to finding the zeros of the function
\begin{eqnarray}\label{equ4}
G(E) & = & E - e \sin E  - M  \\
&&+  \frac{\epsilon^*}{ \left(1-e^2\right)^3} \left[2 \left(e^2+2\right) E - 8 e \sin E +e^2 \sin 2 E\right],\nonumber
\end{eqnarray}

\noindent
for fixed  values of $e$ and $M$. This function is continuous and differentiable on $\mathbb{R}$ for each $(e,M) \in [0,1)\times  \mathbb{R}$. Moreover,  the perturbed Kepler's equation, as well as the classical Kepler's equation, is symmetric with respect to the line of apsides. However, the solutions obtained in the interval $[0,\pi]$ cannot be extended to other ranges of $E$ and $M$ because the values of $G$ are different when $E$, $M$  are replaced with $E\pm 2k\pi$, $M\pm 2k\pi$,  respectively. However, $G$ is a $2\pi$-periodic function only in the eccentric anomaly $E$ for those values of the eccentricity that satisfy the relation
\[e_p=\sqrt{1+\frac{2 \epsilon^* }{3^{1/3} \mathcal{R}}+\frac{\mathcal{R}}{3^{2/3}}},\]

\noindent
where $\mathcal{R}=\sqrt[3]{27 \epsilon^* + \sqrt{3} \sqrt{243 {\epsilon^*}^2-8 {\epsilon^*}^3}}$. Taking  Eq. \eqref{equ4} into account,  
Fig. \ref{fig:fig4} (a) shows a graphical representation of $e_p(a,i)$. The units of $a$ and $i$ are   mean equatorial planet  radii and radian, respectively.  
$e_p$ only exists  for $i\in [0,r_1)  \cup(r_2,\pi]$; when  $a=\alpha$, the roots  of the equation $e_p=1$ are  $r_1=\arcsin(\sqrt{2/3})$ and $r_2=\pi -\arcsin(\sqrt{2/3})$ (Fig. \ref{fig:fig4} (b)). Remember that the classical KE is recovered for  the values of  inclination $r_1$ and $r_2$.  On the other hand, the value of $e_p$ increases when  $a$ increases.

\begin{figure}
\begin{subfigmatrix}{1}
\subfigure[$G$ plots for $i_1=0$ rad,  that is, $\epsilon^*<0$,  $M=0$ rad  and several high values of  eccentricity.]{\includegraphics{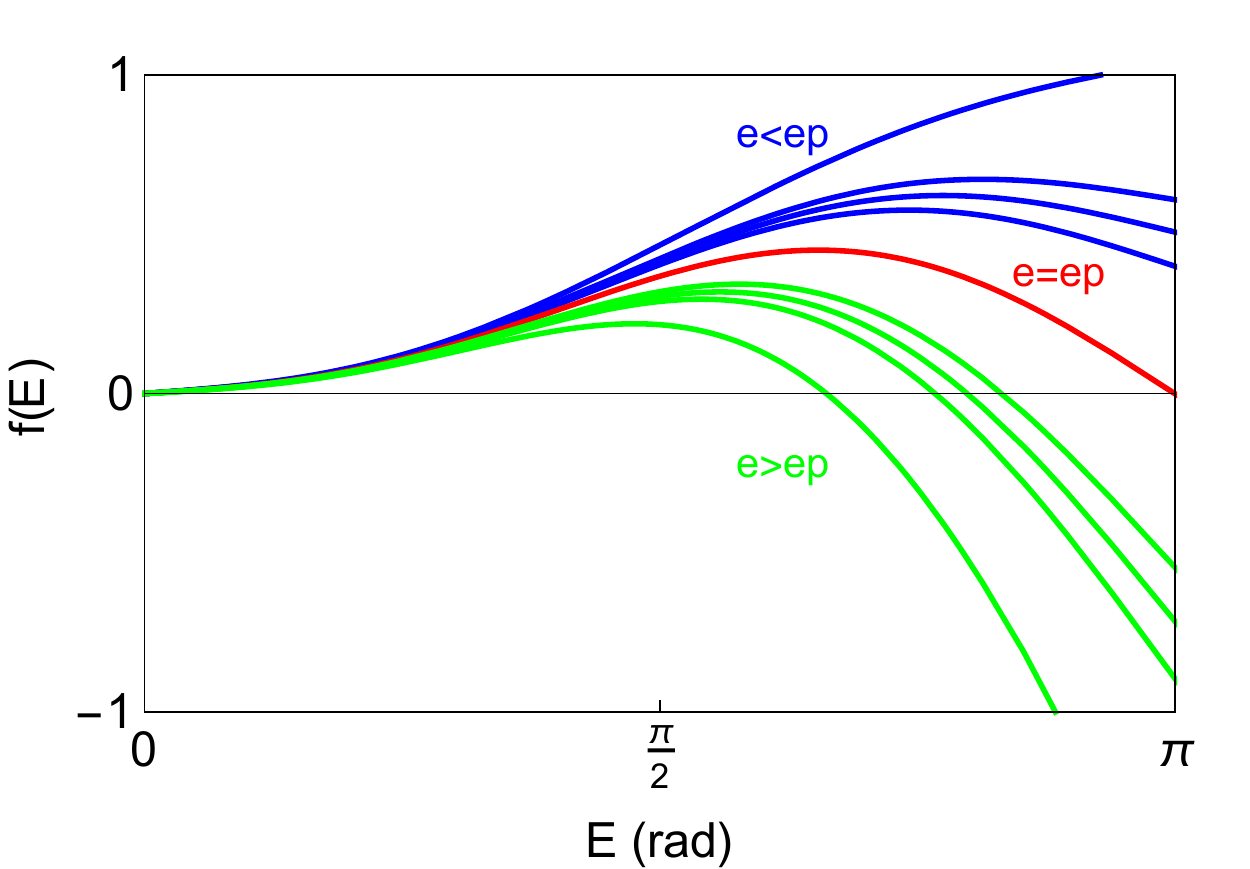}}
\subfigure[Solutions of $G$ for $M=\pi$ and $0\leq e < 1$. ]{\includegraphics{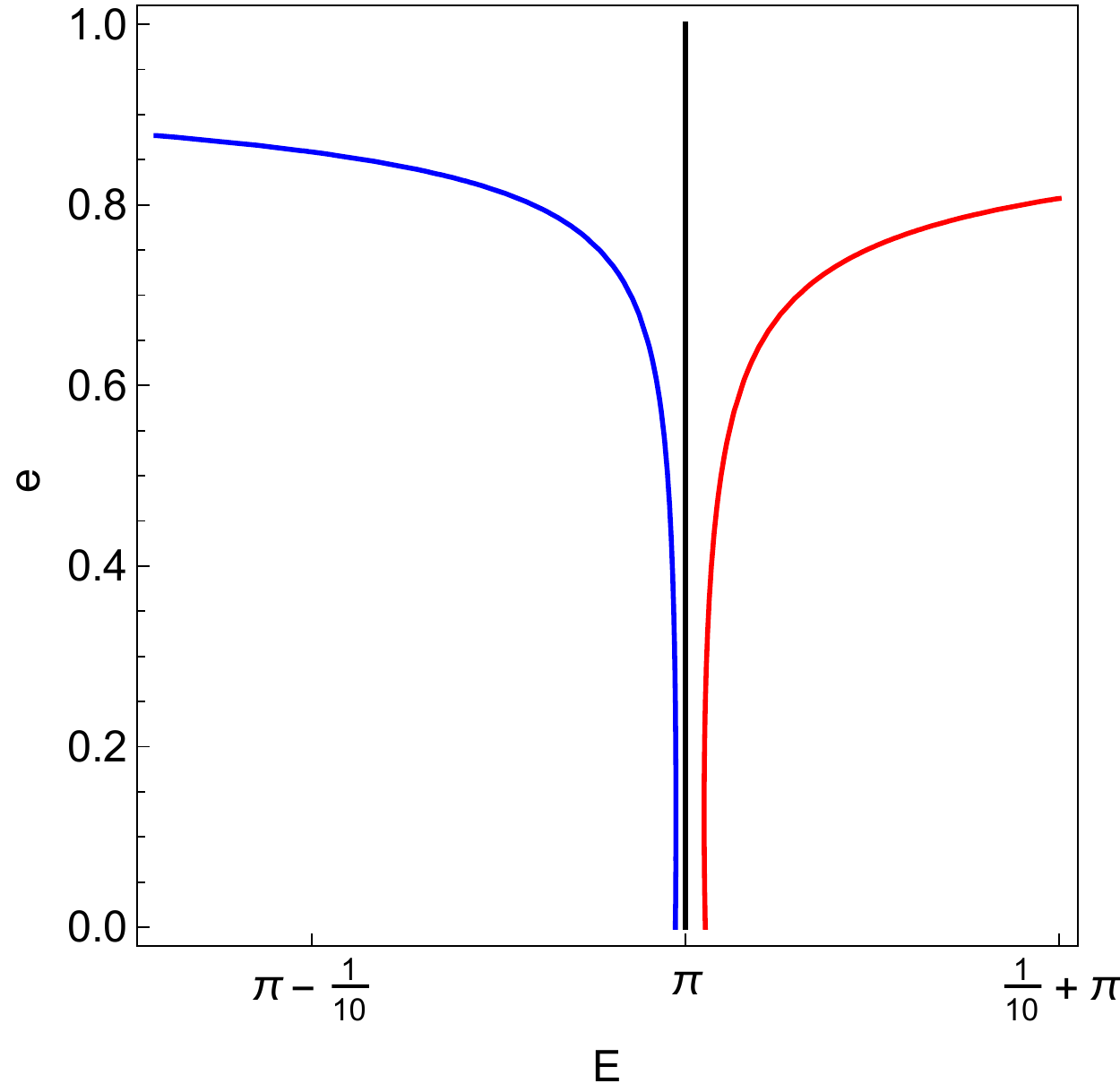}}
 \end{subfigmatrix}
 \caption{Graphical representations for  $a= 7200$. }
 \label{fig:fig6}
\end{figure}

In general, the solution of  $G(E)=0$ is not unique in the interval $[0,\pi]$. In particular, when $i\in [0,r_1)  \cup(r_2,\pi]$, that is, $\epsilon^*<0$, and $e\geq e_p$, the number of solutions   are two, whereas for $e<e_p$ we only have one solution, as can be seen in Fig. \ref{fig:fig6} (a). In the particular case of $e=e_p$, these solutions  are $E=0$  and $E=\pi$.  
On the other hand,  when $i\in [r_1,r_2]$, that is, $\epsilon^*\geq0$, the function $G$ is monotone, and then the solution is unique for $0\leq e < 1$. Finally,   it is not possible to guarantee that for $M \in [0,\pi]$ then $E \in [0,\pi]$. Fig. \ref{fig:fig6} (b) shows all the  solutions of the equation for  $M=\pi$ and $a= 7200$ km. The red line shows all the solutions when $\epsilon^*<0$; as can be seen, the solutions are out of the interval $[0,\pi]$. For high values of the eccentricity, the value of the solution increases, that is, the solution moves away from the interval. The black line, which corresponds to $\epsilon^*=0$, represents the classical KE, in which case whenever  $M=\pi$, the solutions are $E=\pi$ for any eccentricity. Finally, the  blue line  corresponds to $\epsilon^*>0$, case in  which all the solutions are contained in the interval $[0,\pi]$. In the case of  $M=0$,   part of the solutions are out of the interval only for negative values of $\epsilon^*$ and $e<e_p$.

\section{Numerical Results}

In this section,  the  Danby method (DM) and three initial guesses  are applied so as to solve the GKE. It is worth noting that, in order to solve Eq. \eqref{equ4} with this iterative method, it is necessary  to calculate the first, second and third derivatives of $G$ with respect to $E$, 
\begin{eqnarray*}
G'(E) & = & 1 - e \cos E  +  \frac{2\epsilon^*}{ \left(1-e^2\right)^3} \left[ \left(e^2+2\right)  
 \right. \\ && \left. - 4 e \cos E + e^2 \cos 2 E\right], \\
G''(E) & = &  e \sin E  +  \frac{4 e \epsilon^*}{ \left(1-e^2\right)^3} \left[ 2  \sin E - e \sin 2 E\right], \\
G'''(E) & = &  e \cos E  +  \frac{8 e \epsilon^*}{ \left(1-e^2\right)^3} \left[   \cos E -  e \cos 2 E\right]. 
\end{eqnarray*}

Table \ref{tab:label} shows the initial guesses used in our study. The first two ones have been proposed in scientific literature for solving KE: $S_1$ is a classical and simple function of M, whereas in   $S_2$  the computation is divided into two regions (see Reference \cite{dan1983_kepeq1} for more details). Finally,  $S_3$ is the solution  of  Kepler's equation itself, which is also calculated using the  Danby method.

\begin{table}
    \caption{ Initial guesses used to solve the GKE.} \label{tab:label}
        \centering 
   \begin{tabular}{c   l l} 
          Id & $E_0$\\
      \hline 
$S_1$ &    $ M$ &\\[1ex]
$S_{2}$ &  $ M + e^2 (\sqrt[3]{6M}-M)$  & if  $M<0.1 $\\[1ex]
 &  $  M + 0.85e$  &if $M \geq 0.1$\\[1ex]    
 $S_3$ &    $ \mathrm{solution\: of \:KE}$&\\[1ex]
   \end{tabular}
\end{table}

Then, this iterative method is used to solve the GKE  for a  grid  of points in the $M$--$e$ plane ($0\leq M \leq \pi$, $0\leq e < 1$), separated by a uniform space of $\Delta M=1/1000$ rad and $\Delta e=1/1000$; the number of points in the grid is therefore $3142000$. It is worth noting that this study is restricted to the interval  $M\in[0, \pi]$ because, in our problem, the GKE appears as a perturbed case of the KE, although  an extensive analysis should be done for all $\mathbb{R}$ from the mathematical point of view.

\begin{figure}
\begin{subfigmatrix}{2}
\subfigure[$S_{1}$.]{\includegraphics{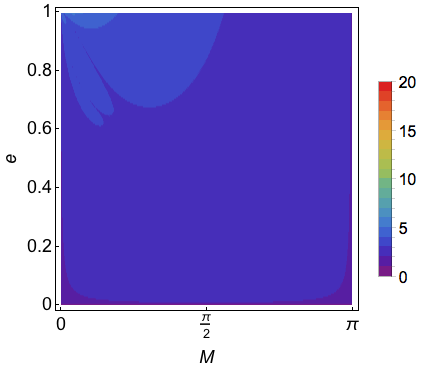}}
\subfigure[$S_{2}$. ]{\includegraphics{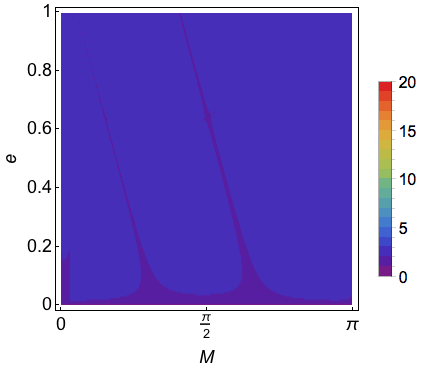}}
 \end{subfigmatrix}
 \caption{The colour scale shows the number of iterations of the Danby method    for the value of  inclination $i_3=r_1$ ($\epsilon^*=0$) using   $S_{1}$ and $S_{2}$  as initial guesses. The   method always converges in both cases.  $S_1$ needs  between 3 and 4 iterations  in  $95.91\%$ of the cases, whereas  $S_2$ only needs between 2 and 3 iterations in $93.36\%$ of the cases.}
 \label{fig:fig8}
\end{figure}

\begin{figure}
\begin{subfigmatrix}{2}
\subfigure[$i_1=0^\circ$. ]{\includegraphics{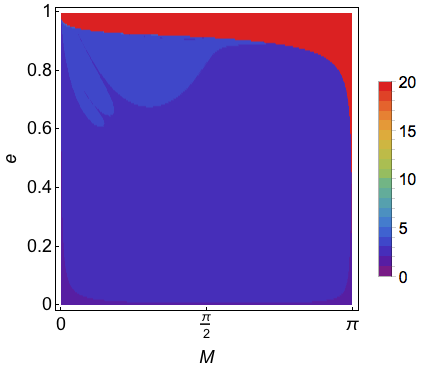}}
\subfigure[$i_2=53^\circ$. ]{\includegraphics{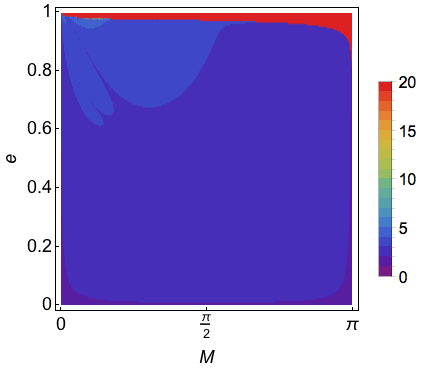}}
\subfigure[$i_4=55^\circ$. ]{\includegraphics{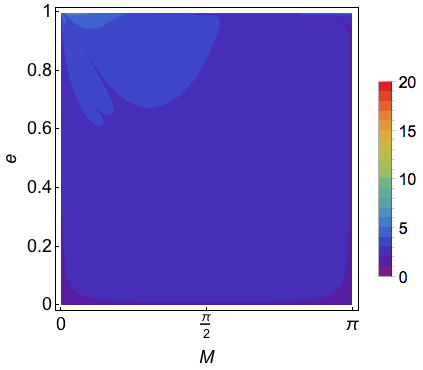}}
\subfigure[$i_5=90^\circ$.]{\includegraphics{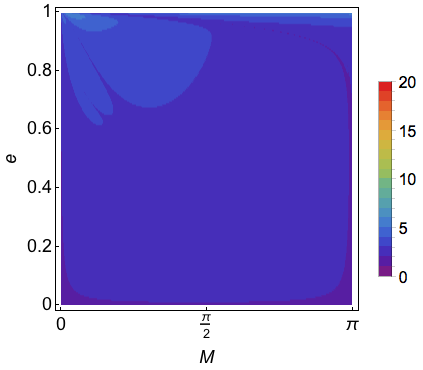}}
\end{subfigmatrix}
\caption{The colour scale shows the number of iterations of the Danby method   using   $S_{1}$ as initial guess.  $\epsilon^*$ takes negative values for $i_1$ and $i_2$, and positive values for  $i_4$ and $i_5$. The red colour represents non-convergent regions. Most of the cases  require between 4 and 5 iterations to converge to the solution  for $i_1$ and $i_2$, with percentages of $87.44\%$  and $93.04\%$  respectively, whereas the number of iterations practically reduces by one  for  $i_4$ and $i_5$, with percentages of  $95.63\%$ and  $94.22\%$, respectively.}
\label{fig:fig9}
\end{figure}

\begin{figure}
\begin{subfigmatrix}{2}
\subfigure[$i_1=0^\circ$.]{\includegraphics{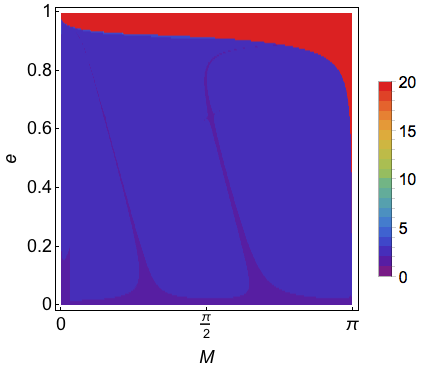}}
\subfigure[$i_2=53^\circ$. ]{\includegraphics{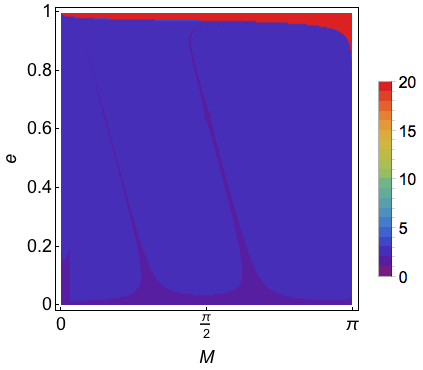}}
\subfigure[$i_4=55^\circ$. ]{\includegraphics{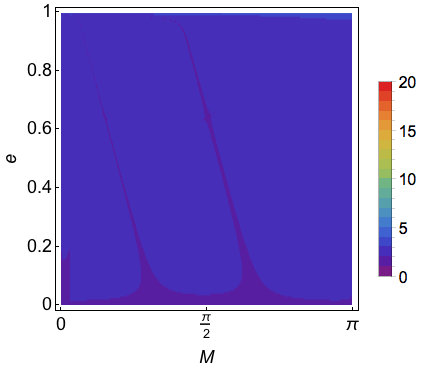}}
\subfigure[$i_5=90^\circ$.]{\includegraphics{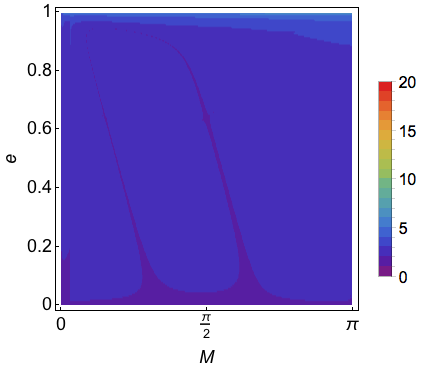}}
\end{subfigmatrix}
\caption{The colour scale shows the number of iterations of the Danby method  using   $S_{2}$ as initial guess. $\epsilon^*$ takes negative values for $i_1$ and $i_2$, and positive values for  $i_4$ and $i_5$. The red colour represents non-convergent regions. Most of the cases require  between 2 and 3 iterations to converge to the solution  for the four inclinations $i_1$, $i_2$, $i_3$ and $i_4$, with percentages of $89.25\%$, $96.35\%$, $99.86\%$ and  $98.63\%$, respectively.}
\label{fig:fig10}
\end{figure}

\begin{figure}
\begin{subfigmatrix}{2}
\subfigure[$i_1=0^\circ$.]{\includegraphics{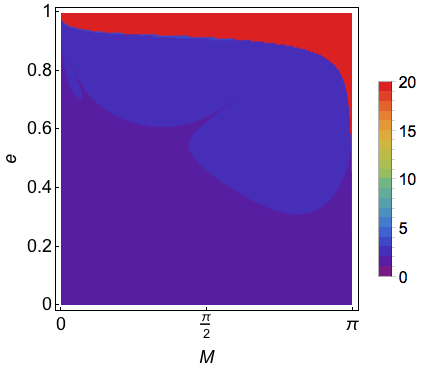}}
\subfigure[$i_2=53^\circ$. ]{\includegraphics{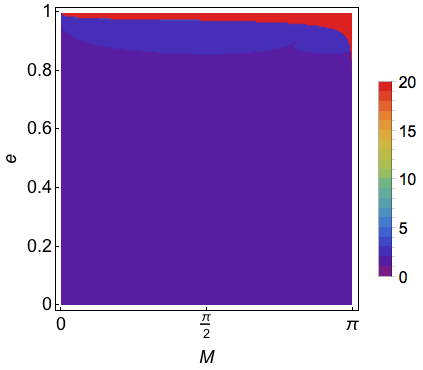}}
\subfigure[$i_4=55^\circ$. ]{\includegraphics{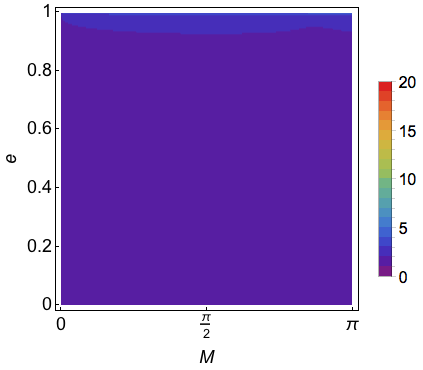}}
\subfigure[$i_5=90^\circ$.]{\includegraphics{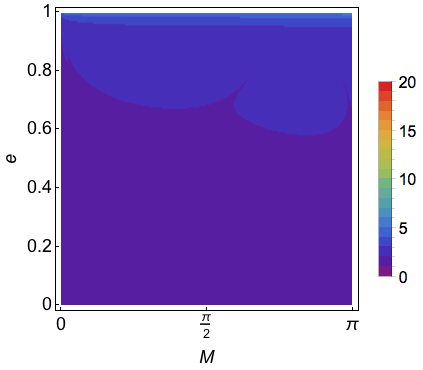}}
\end{subfigmatrix}
\caption{The colour scale shows the number of iterations of the Danby method  using   $S_{3}$ as initial guess. $\epsilon^*$ takes negative values for $i_1$ and $i_2$ and positive values for  $i_4$ and $i_5$. The red colour represents non-convergent regions. The number of cases that require between  2 and 3 iterations to converge to the solution represent $50.27\%$,  $86.87\%$, $93.38\%$ and $66.85\%$  for the  inclinations $i_1$, $i_2$, $i_3$ and $i_4$, respectively.}
\label{fig:fig11}
\end{figure}

The maximum number of iterations allowed is 20, and the convergence is considered to be achieved if  $\|E_{i+1}-E_i\| \leq10^{-14}$.  The selected planet  is the Earth, for which  five  different inclinations are compared. The  first two values, $i_1=0^\circ$ and $i_2=53^\circ$, correspond to negative values of $\epsilon^*$, the third value is  $r_1$, where the GKE is reduced to KE, and  the  last two values are $i_4=55^\circ$ and $i_5=90^\circ$, which correspond to positive values of $\epsilon^*$.  On the other hand, several semi-major axes have been considered in this study, although in the following   discussion the semi-major axis  has been set to $7200$ km, which corresponds to a LEO orbit. Finally,  an additional  convergence criterion is required to determine the relation between the generalized anomalies $E$ and $M$; the iterative method converges if  the root of  Eq. \eqref{equ4} belongs to the interval $[0,\pi]$. The results  of this study are summarized in  figures \ref{fig:fig8}-\ref{fig:fig11}, in which  the $M$--$e$ plane  has been divided into regions that correspond to the same  number of iterations during the resolution of the equation. The number of points plotted  in each graph is 628400.

Fig. \ref{fig:fig8}  shows the results of the application of  DM with the initial guesses $S_1$ and $S_{2}$ in the case in which  GKE is reduced to KE, that is, for the value of the inclination $i_3=r_1$. In both cases, the   method always converges and  needs  between 3 and 4 iterations  for  $95.91\%$ of the cases when   $S_1$ is used, and between 2 and 3 iterations ($93.36\%$) for  $S_2$. For more details regarding the full analysis of this combination, see References \cite{dan1983_kepeq1,dan1987_kepeq3}. 

Fig. \ref{fig:fig9} shows the results of the application of    DM with the initial guess $S_1$.   The case $\epsilon^*<0$ is illustrated in Figs. \ref{fig:fig9}  (a) and (b).  Most of the cases  converge to the solution using between 4 and 5 iterations, with percentages of $87.44\%$
for  $i_1=0^\circ$  and  $93.04\%$ for $i_2=53^\circ$. It is worth noting that there are  non-convergent regions (red colour in Figures); these regions reach their maximum size ($10.48\%$ of the cases) for $i_1=0^\circ$, and decrease as the inclination increases, until  they disappear for $i_3=r_1$. For $i_2=53^\circ$ the non-convergent region represents  $3.58\%$ of the  cases, and corresponds to values  $(e,M)$   that cause GKE to have   two solutions or one solution  out of the interval $[0,\pi]$.   Figs. \ref{fig:fig9}  (c) and (d) correspond to  $\epsilon^*>0$:  the cases that converge to the solution using between 3 and 4 iterations are  $95.63\%$ for $i_4=55^\circ$ and  $94.22\%$ for $i_5=90^\circ$.  It is worth noting that the method  has convergence problems for very high eccentricities $e>0.99$. The change in the shape of the  regions in the $M$--$e$ plane with respect to the KE case can be seen  for high eccentricities ($e>0.9$). 

The results of the application of    DM with the initial guess $S_2$ are given in  Fig. \ref{fig:fig10} and Table \ref{tab:label2}.    The case $\epsilon^*<0$ is shown in Figs \ref{fig:fig10}  (a) and (b).   The size of the non-convergent regions is  the same as in the cases corresponding to the use of $S_1$ as the initial guess.   DM  uses between 2 and 3 iterations in  $89.25\%$ and  $96.35\%$ of the cases for $i_1=0^\circ$  and $i_2=53^\circ$, respectively.  Finally, Figs \ref{fig:fig10} (c) and (d) analyse the case $\epsilon^*>0$.    DM  uses between 2 and 3 iterations in  $99.86\%$ and  $98.63\%$ of the cases for $i_4=55^\circ$  and $i_5=90^\circ$, respectively. The method  also has convergence problems for very high eccentricities, $e>0.99$.
The change in the shape of the  regions in the $M$--$e$ plane with respect to the KE case is also present   for high eccentricities ($e>0.9$).

\begin{table}
   \caption{Percentage of cases that require 2 or 3 iterations, when the Danby method is applied to the inclinations $0^\circ$, $53^\circ$, $55^\circ$  and $90^\circ$, for both S2 and S3 as the initial guess.}
   \label{tab:label2}
        \centering 
   \begin{tabular}{c |c |c | c  } 
Inclination & Initial Guess & 2 iterations & 3 iterations\\ 
       \hline 
$i_1=0^\circ$ & $S_2$  & $\phantom{0}6.31\%$ & $82.94\%$\\[.5ex]
& $S_3$  &  $50.27\%$ & $38.98\%$\\[1.5ex]
$i_2=53^\circ$ & $S_2$& $\phantom{0}6.57\%$ & $89.78\%$\\[.5ex]
& $S_3$  &  $86.87\%$ & $\phantom{0}9.47\%$\\[1.5ex]
$i_3=55^\circ$ & $S_2$&  $\phantom{0}6.50\%$ & $93.36\%$\\[.5ex]
& $S_3$  &  $93.38\%$ & $\phantom{0}5.54\%$\\[1.5ex]
$i_4=90^\circ$ & $S_2$& $\phantom{0}6.58\%$ & $92.05\%$\\[.5ex]
& $S_3$ &  $66.85\%$ & $28.48\%$\\
   \end{tabular}
\end{table}

To conclude this  study, the solution of  KE, $S_3$,  is used as the initial guess for the Danby method.   It is an intuitive initial guess due to the fact that GKE can be considered as a perturbed version of  KE.  The results are given in  Fig. \ref{fig:fig11}.   The non-convergent regions for 
the case $\epsilon^*<0$ represent  the same percentages  as when $S_1$ and $S_2$ are taken as initial guesses.  $S_3$ achieves convergence to the solution in only 2 iterations in more than $50\%$ of the cases, as can be seen in  Table \ref{tab:label2};  this percentage increases up to $85$--$100\%$ when the inclination takes values close to $r_1$,  for which the GKE is reduced to KE. In summary,
DM  requires between 2 and 3 iterations in  $89.25\%$, $96.34\%$, $98.92\%$ and  $95.33\%$ of the  cases for $i_1=0^\circ$, $i_2=53^\circ$, $i_4=55^\circ$ and $i_5=90^\circ$, respectively.     DM also has convergence problems  for $e>0.99$ in the case $\epsilon^*>0$.  

As it has already been mentioned in the previous section, the value of $|\epsilon^*|$ decreases, and, therefore also its influence, when  $a$ increases. 
This implies that the size of the non-convergent regions  will be smaller, and  the shape of the  $M$--$e$ plane solutions for GKE will be  similar to KE for high eccentricities. 

\begin{figure}
\centering\includegraphics[width=6cm]{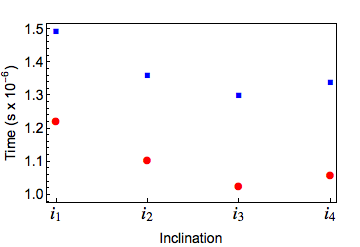}
\caption{Computational time required to reach an error of $10^{-14}$ with the Danby method. Red circle and blue square represent the initial guesses $S_2$ and $S_3$, respectively.} 
\label{fig:fig12}
\end{figure}

The best  computational time needed to achieve a convergence error of $10^{-14}$  with the Danby method is obtained with $S_2$ (red circle), whereas the worst is reached with $S_3$ (blue square), as can be seen in Fig \ref{fig:fig12}.  In particular, $S_2$ is approximately $20\%$ faster than $S_3$ for all the inclinations considered in this study.  The CPU used has been an Intel Core i7 with a clock frequency of  1.7 GHz. 

\section{Conclusion}

In this work, a first approach to the problem of solving the generalized Kepler's equation by using an iterative  method proposed by Danby and Burkardt, together with  two habitual initial guesses \cite{dan1983_kepeq1} used to solve Kepler's equation, $S_1$ and $S_2$, and even with the solution  of KE itself as an initial guess for GKE, $S_3$, has been tested.  At first order, GKE is a function of the eccentricity, the mean and eccentric anomalies, and a small parameter, $\epsilon^*$, which depends on the semi-major axis, the inclination and the  physical parameters $\alpha$ and  $J_2$. The value of the small parameter $\epsilon^*$ can be negative, zero or positive: its sign is a function of the inclination, whereas its magnitude $|\epsilon^*|$ decreases when  $a$ increases. On the other hand, when $\epsilon^*=0$,  GKE is reduced to  KE.

For the initial guesses $S_1$ and $S_2$, the behaviour of  the  iterative method, when  GKE is solved, is similar  to the KE case for $e<0.9$. For high eccentricities, $e>0.9$, the behaviour changes and  non-convergent regions  appear for $\epsilon^*<0$; in these regions we can simultaneously find  convergence problems of the iterative method,  multiple solutions of    GKE for a value of the eccentricity $e_p$, and solutions that are not contained in the interval $[0,\pi]$, property that  KE verifies. On the other hand, $S_2$ and $S_3$  achieve convergence to the solution in only 2 or 3 iterations in more than   $90\%$ of the cases, increasing the number of iterations when the eccentricity grows. However, $S_3$  only needs 2 iterations to achieve convergence to the solution in more than $50\%$ of the cases;  this percentage increases up to $85$--$100\%$ when the inclination takes values close to $r_1$ for which the GKE is reduced to KE. It is worth noting that the decrease of the number of iterations is at the expense of speed, $S_3$ is approximately $20\%$ slower than $S_2$ for all the inclinations considered in this study.

\section*{Acknowledgements}

This work has been funded by the Spanish State Research Agency and the European Regional Development Fund under Project ESP2016-76585-R (AEI/ERDF, EU).  The authors would like to thank an anonymous reviewer for his/her valuable suggestions.




\bibliographystyle{mnras}
\bibliography{GKEquation} 




\appendix

\section{Generating function}

The first-order generating function of the elimination of the parallax is given by

\begin{eqnarray*}
\mathcal{W} &=&  \left(\Frac{\alpha}{p'}\right)^{2} \Frac{\Theta'}{8}\left[ S' (4 - 9 s'^2)  \cos \theta'
+S' s'^2 \cos 3 \theta' \right.\\[1.ex] 
&&\qquad\qquad\left. \phantom{-} 
+ C' (-4 + 3 s'^2)  \sin \theta' -3 s'^2 \sin 2 \theta' \right.\nonumber\\[1.ex] 
&&\qquad\qquad\left. \phantom{-} - C' s'^2 \sin 3\theta'
 \right],\nonumber
 \end{eqnarray*}

\noindent
where $p'=\Theta'^2/\mu$, and $C'$, $S'$ are given as functions of polar-nodal variables by  

\begin{eqnarray*}
C' & = & \left(\frac{p'}{r'}-1\right) \cos  \theta' + \frac{p'R'}{\Theta'}\sin \theta', \\[.5ex]
R' & = & \left(\frac{p'}{r'}-1\right)\sin \theta'-\frac{p'R'}{\Theta'}\cos \theta'.
 \end{eqnarray*}


\bsp	
\label{lastpage}
\end{document}